\documentclass[12pt,preprint]{aastex}
\shorttitle{Long-Term Optical Variability of NGC 2992}
\shortauthors{Trippe and Crenshaw}
\usepackage{natbib}
\begin{document}
\title{Long-Term Variability in the Optical Spectrum of the Seyfert Galaxy NGC 2992 }
\author{M. L. Trippe, D. M. Crenshaw, and R. Deo\altaffilmark{1} }
\affil{Department of Physics and Astronomy, Georgia State University, One Park Place South SE, Ste. 700, Atlanta, GA 30303}
\email{trippe@chara.gsu.edu, crenshaw@chara.gsu.edu}
\author{M. Dietrich}
\affil{Department of Astronomy, The Ohio State University, 4055 McPherson Lab, 140 W. 18th Ave., Columbus, OH 43210}
\email{dietrich@astronomy.ohio-state.edu}
\altaffiltext{1}{currently at the Drexel University Department of Physics, Disque Hall, South 32nd St., Rm. 813, Philadelphia, PA 19104; email: rpd@physics.drexel.edu}
\begin{abstract}
New spectra of NGC 2992 from the Cerro Tololo Inter-American Observatory show that this nearby AGN has changed its type classification to a Seyfert 2 in 2006. It was originally classified as a Seyfert 1.9, and has been previously seen as a Seyfert 1.5 with strong broad H$\alpha$ emission. A comparison of the reddening and equivalent hydrogen column density derived for the narrow-line region from these new data with those previously calculated for different regions closer to the nucleus shows them to be very similar, and suggests that these different regions are all being absorbed by the same opacity source, a large 100-pc scale dust lane running across the nucleus. However, obscuration by dust in this lane is probably not responsible for classification changes which occur in only a few years. It is more likely that NGC 2992's observed variations are due to a highly variable ionizing continuum. We therefore conclude that, although NGC 2992 was originally identified as a Seyfert 1.9, this was not because of an oblique viewing angle through the atmosphere of a central dusty torus, but because its active nucleus was identified when it was in a low continuum state.
\end{abstract}

\section{Introduction}
     Seyfert galaxies are a nearby low-luminosity type of active galactic nucleus (AGN). They are customarily divided into two types, based on their emission-line spectra, following the scheme of \citet{kha71}. Those Seyferts whose spectra show both broad and narrow components of permitted emission lines, as well as narrow forbidden lines, are classified as Type 1. Those whose spectra manifest only narrow emission lines are classified as Type 2. Furthermore, the spectra of Type 2 contrast with Type 1 in that they exhibit only a weak, mostly stellar continuum; the spectra of Type 1 Seyferts are characterized by a strong non-stellar continuum. According to the unified model of AGN (see Antonucci 1993 for review), the differences observed between these two types are due to differences in the AGN's orientation with respect to the observer's line of sight. This theory asserts there to be a torus-shaped region of dusty molecular gas surrounding the innermost region of the nucleus, i.e., the area where the broad emission lines originate around the non-stellar continuum source.  Seyfert 1's, then, have an orientation that allows a clear view of the central region, while Seyfert 2's have an orientation such that the material in this torus intercepts our line of sight and obstructs our view of the broad emission-line region (BLR) and central continuum source.

     \citet{ost81} added two important intermediate classifications to this simple dichotomic classification system. Seyfert 1.8 galaxies show weak broad H$\alpha$ and H$\beta$, while Seyfert 1.9 galaxies show just weak broad H$\alpha$. These AGN also show weaker featureless continua and larger broad Balmer decrements than galaxies classified as Type 1. Osterbrock initially suggested the observed characteristics of Seyfert 1.8s and 1.9s could be due to the reddening of the continuum and BLR by dust. In terms of the unified model, this would indicate a line of sight that grazes the outer edge of the obscuring dusty torus. However, it was later seen from theoretical models that these observed characteristics could alternatively be produced without dusty material external to the broad-line clouds, if the clouds themselves have fairly low optical depths and ionization parameters \citep{rud83}. This is known as the ``($\tau$, U)'' theory. There is evidence in support of both these hypotheses in different objects \citep{goo95}, which intimates that the intermediate Seyfert classes may encompass two different physical cases.

      One simple way to assess which physical phenomenon is responsible for a Seyfert's intermediate-type appearance is to observe it over time and note any changes in its broad components. If the timescale of variation is too short to be consistent with reddening due to the motion of dusty material across the line of sight to the continuum, we must revert to the low ionization parameter/low ionizing flux theory. This motivation interested us in the optical monitoring of Seyfert 1.8s and 1.9s over time. 

      NGC 2992 is a nearby galaxy with redshift z=0.00771 that has been well studied in all wavelength regimes. Long-slit observations have demonstrated the complex kinematics of the ionized gas in its circumnuclear region \citep{hec81, col87}. Nuclear optical spectra show it to be a Seyfert galaxy, but its type classification has been observed to vary conspicuously in the past, leading to classifications ranging from Seyfert 1 to Seyfert 2. Early spectra published by \citet{shu80}, \citet{ver80}, and \citet{war80} all show the presence of a weak broad H$\alpha$ component but no detectable corresponding broad H$\beta$ component, leading to its original classification as a Seyfert 1.9. When it was observed in 1994 by \citet{all99}, however, it had apparently lost its broad H$\alpha$ component and was classified as a Seyfert 2. A 1999 spectrum taken by Gilli et al. at the ESO NTT showed it to have regained its broad H$\alpha$ emission \citep{gil00}. Gilli et al. also correlated the presence of broad H$\alpha$ with the galaxy's X-ray flux; they found NGC 2992 to have been in a high X-ray state when it was initially observed circa 1980, but that the X-ray flux had been slowly decreasing over time and was at a minimum when Allen et al. observed its optical spectrum in 1994. At the time of Gilli et al.'s observation of broad H$\alpha$ in 1999, they used BeppoSAX data to show that NGC 2992 had returned again to its previous active X-ray state, and postulated these variations were due to different phases of rebuilding of the central accretion disk. The interesting history of variation of this object, as well as its intermediate type classification, induced us to choose it as the subject of an optical monitoring campaign. 

\section{Observations and Data Reduction}
	We monitored NGC 2992 for about a year and a half, from 2006 January to 2007 June, using the R-C spectrograph on the Cerro Tololo Inter-American Observatory (CTIO) 1.5-m telescope in Chile. We obtained observations on a monthly basis, with the exception of when the galaxy was too near the Sun to be observed. Spectra were observed using two different settings: one with a grating with a resolution of 4.3 \AA\ (giving a dispersion of $\sim$1.47 \AA\ $/$pixel) to take blue spectra from approximately 3660-5440 \AA\ to include H$\beta$, and the other with a resolution of 3.1 \AA\ $($$\sim$1.10 \AA\ $/$pixel dispersion) and a Schott GG 495 filter to take red spectra from approximately 5650-6970 \AA\ to include the H$\alpha$ line.  We used a slit at a 90$^{\circ}$ P.A. centered on the galaxy's nucleus to obtain accurate absolute fluxes on photometric nights. The majority of the spectra were taken using a 2'' slit width, but observations were also taken on nights when other slit widths were in use. Table 1 chronicles the dates and settings of our observations. To eliminate cosmic ray hits, we took three exposures each time the galaxy was observed. The stars LTT 4364 and Feige 110 were also observed with these setting for the purpose of flux calibration. The spectra were then reduced and flux calibrated using standard IRAF reduction packages for long-slit spectroscopy. Since the spectra were seen to have remained essentially constant with time, we averaged them to obtain a higher S/N ratio in our final resulting spectrum. As shown in Figure 1, the final spectrum looks like that of a Seyfert 2, in that only narrow emission lines are present and it exhibits strong stellar absorption features. 

	In order to assess the impact of using slits of different sizes, we compared the integrated fluxes of the [S II] $\lambda$6716 and $\lambda$6731 lines in a spectrum taken with the 2'' slit width with the integrated fluxes of the lines in a spectrum taken with the 4'' slit width. We found there to be only about a 20\% difference between the two. The final spectrum we present is an average of all 10 observations, only 4 of which were made with slits larger than 2''. We therefore estimate the fluxes of these narrow lines to be increased by only (4/10)*22\%= 8\% over what they would be if a 2'' slit width had been used exclusively throughout the observations.

	To obtain a spectrum suitable for measurement of nuclear emission lines, the light from the nucleus must be isolated from that of the host galaxy. We removed the host galaxy spectrum from the CTIO data by subtracting off a normal galaxy spectrum \citep[from][]{kin96} scaled to give the optimum fit to the observed average spectrum's continuum and absorption line features. The remaining continuum was fit by a power-law with $\alpha=-1.83$ and subtracted off, but with little effect as this component was very weak, consistent with the rest of the evidence that NGC 2992 was a Type 2 Seyfert at the time of the CTIO observations. Figure 2 shows the resulting spectrum after these subtractions.

\section{Data Analysis}
 	We determined the reddening of the narrow-line region from the narrow components of H$\alpha$ and H$\beta$, assuming their intrinsic ratio to be equal to the recombination value of $H\alpha/H\beta=2.9$ \citep{ost06}, and further assuming the standard Galactic reddening curve of \citet{sav79} to be applicable. By this method we found $E(B-V)=0.71\pm0.11$ mag, and de-reddened the spectrum accordingly. We give measured values of NGC 2992's observed and de-reddened line ratios relative to H$\beta$ in Table 2. 

	To study the H$\alpha$ profile, we deblended the narrow [N II] $\lambda\lambda$6548, 6583 \AA\ lines flanking each side of H$\alpha$ by using the distinctly narrow component of the H$\alpha$ line as a template for each of the doublet lines in an iterative process. The template was moved to the position of each [N II] line, scaled in width and height to match it, and subtracted off. The height of the line at 6583 \AA\ was set to three times that of the line at 6548 \AA\ based on the ratio of their transition probabilities. The profile before and after subtraction of the [N II] lines is shown in Figure 3. If there is a broad component, it is certainly very faint, no more than 30\% of the flux of the narrow component. 
	
\section{Results}
      Given its history of extreme variability, NGC 2992 has remained at a remarkably constant low state over the year and a half of our observations. As can be seen from Figure 4 where three representative spectra (from the beginning, middle and end of the campaign) are plotted, NGC 2992 showed little, if any, variation over the observation period. The H$\alpha$ profile showed no sign of change in any broad component. Furthermore, the continuum did not change over this time period, in line with our conclusion that it was dominated by stellar light. 

      The observations, when compared with the historical spectra mentioned in the introduction, do at least prove that this Seyfert has undergone yet another dramatic change, from the high flux state it was in when last observed by Gilli et al. in 1999, back to the low state observed by Allen et al. in 1994. Figure 5 plots the data from Figure 8 in \citet{gil00}, three spectra all observed with 2'' slit widths taken over the course of 31 years, along with our new average spectrum around H$\alpha$. It can be seen that NGC 2992 lost its broad H$\alpha$ in 1994, recouped it again by 1999 \citep[as noted by][]{gil00}, but has now lost it again in our spectra.

      By using the observed reddening of the narrow-line region, $E(B-V)=0.71$ mag, and the relation $N_{H}=5.2\times10^{21}E(B-V)$ cm$^{-2}$ derived by \citet{shu85} for the local ISM, we estimate the column density absorbing the narrow-line region to be $N_{H}=3.7\pm0.6\times10^{21}$ cm$^{-2}$. This is equivalent, within the error estimates, to the value derived by Gilli et al. from the ratio of narrow $Pa\beta/H\alpha$ of $E(B-V)$ of 0.65$\pm$0.19 mag ($N_{H}=3.4\pm1.0\times10^{21}$ cm$^{-2}$). Gilli et al. also found $E(B-V)=0.71\pm$0.19 mag ($N_{H}=3.7\pm1.0\times10^{21}$ cm$^{-2}$) for the BLR in 1999, using a near IR spectrum exhibiting broad Br$\gamma$ and Pa$\beta$ lines. The nearly identical reddenings of the BLR and NLR indicate a dust screen external to the NLR, as suggested by \citet{gil00}. 

      X-ray observations have found absorption values similar to those observed in the optical. \citet{yaq07} found $N_{H}$ to be $7.99\times10^{21}$ cm$^{-2}$ from Suzaku observations in November and December of 2005, only a month before we began our optical monitoring at CTIO. This value is similar to Gilli et al.'s measurement with BeppoSAX of $N_{H}=9.0\pm0.3\times10^{21}$ cm$^{-2}$, from when the galaxy was in a high X-ray and optical state. Throughout the history of NGC 2992's observations, the X-ray flux has been seen to vary dramatically, from $f_{2-10keV}=0.63\times10^{-11}$ ergs cm$^{-2}$ s$^{-1}$ observed by Gilli et al. with BeppoSAX in 1997 \citep{gil00} to $f_{2-10keV}=8.88\times10^{-11}$ ergs cm$^{-2}$ s$^{-1}$ observed by Murphy et al. with RXTE in 2005 \citep{mur07}, but the X-ray column has remained virtually constant at $\sim10^{22}$ cm$^{-2}$ \citep{gil00,col05,yaq07}. The observed X-ray columns are two to three times higher than those estimated by the reddening values. This may be due to a lower dust-to-gas ratio than that of our Galaxy, or to a separate dust-free X-ray absorber close to the nucleus. 

      To investigate the mid-IR properties of NGC 2992, we retrieved and processed its low-resolution Spitzer IRS spectrum from the Spitzer archives (see Deo et al. 2007 for details). As shown in Figure 6, the spectrum exhibits PAH emission features typical of a strong starburst contribution as well as highly ionized narrow lines, such as [O IV] $\lambda$25.89 $\mu$m and [Ne V] $\lambda$14.31 $\mu$m, attributable to the hard ionizing continuum of the central AGN. The spectrum also exhibits a strong silicate 9.7 $\mu$m absorption feature, which in Seyfert galaxies is which is typically found in Seyfert galaxies with highly inclined ($b/a$ $<$ 0.5) host galaxy disks \citep{deo07}. Indeed, the galactic disk of NGC 2992 is highly inclined, with $b/a=0.31$, according to NED (the NASA/IPAC Extragalactic Database). From the Spitzer spectrum, the trough of the silicate 9.7 $\mu$m feature is at an optical depth of $\tau=0.35$, which corresponds to a reddening of $E(B-V)=2.1$ based on the relationship for diffuse ISM clouds ($A_{V}=18.5\tau_{9.7\mu m}$) given by \citet{roc84}. There are a couple of explanations for this reddening being somewhat higher than the direct values from the BLR and NLR. One possibility is a strong contaminating contribution to the silicate feature from a starburst heavily enshrouded in dust. Another possibility is that the dust may be richer in silicates than Galactic dust, as suggested for the damped Ly$\alpha$ absorber toward AO 0235+164 \citep{kul07}. 
  
\section{Discussion and Conclusions}
      In Table 3, we give a summary of determinations of reddening and/or hydrogen column densities in the line of sight to different components of the AGN.  The fact that these column densities derived from different regions all fall within a about a factor of three of each other suggests that they are all being absorbed by the same opacity source, likely a large-scale dust lane in the host galaxy. An optical Hubble WFPC2 image of the nucleus of NGC 2992 obtained with the F606W filter substantiates this explanation, as it clearly shows a pronounced hundred-pc scale lane of dust passing across the point-source nucleus (see Figure 7). 

      Variable reddening due to dust in this lane is unlikely to be the cause of NGC 2992's history of broad-line variation. A simple calculation of the timescale of variation due to dust in the host galaxy moving at 300 km s$^{-1}$ across a BLR 10 light-days in size gives a timescale $\tau\approx30$ years, much too long to be the cause of variations observed to occur in only a few years. The strong correlation between the flux of the broad component of H$\alpha$ and the 2-10 keV X-ray flux shown by Gilli et al. (2000), as well as the fact that the line of sight column density has remained virtually constant despite these variations, are further indications of an intrinsic origin of NGC 2992's variability. 

      We conclude that NGC 2992 was identified as a Seyfert 1.9 based on its discovery in a low continuum state. It is clear that it is not a case of an oblique line of sight through the atmosphere of a dusty torus; the observed reddening is explained by an external dust lane. This case study of NGC 2992 draws our attention to the question: how many other Seyferts originally identified as Seyfert 1.9s are also not the result of obscuration? Since NGC 2992 is not singular in its variation, as even the prototypical Seyfert 1 galaxy NGC 4151 has been observed to lose and regain its broad lines \citep{kra06}, its history of type fluctuation highlights even broader questions. How often do Seyferts in general change type due to dramatic continuum variations? And, of even more fundamental physical importance, how do these AGN manage to become completely quiet on timescales of about one year? To answer these important questions and sort out the apparently diverse category of Seyfert 1.8/1.9s, more of these objects need to be monitored on a long term (yearly or longer) basis. We will continue observing NGC 2992 in the optical to determine the exact timescale of variation in its broad lines. 

\acknowledgments
We would like to thank Robert Maiolino for kindly providing us with the optical spectra used for comparison in Figure 5. Our spectra of NGC 2992 were taken at the CTIO 1.5-m telescope, operated by the SMARTS Consortium. We would like to thank Todd Henry for helping to provide access to the SMARTS time. This work has made use of the NASA/IPAC Extragalactic Database (NED), which is operated by the Jet Propulsion Laboratory, California Institute of Technology, under contract with NASA. The spectrum in Figure 6 was retrieved from archival data obtained with the $Spitzer$ $Space$ $Telescope$, which is also operated by the Jet Propulsion Laboratory, California Institute of Technology, under a contract with NASA. The image of NGC 2992 in Figure 7 was obtained from the Multimission Archive at the Space Telescope Science Institute (MAST). STScI is operated by the Association of Universities for Research in Astronomy, Inc., under contract with NASA's Goddard Space Flight Center, Greenbelt, MD. 

\bibliographystyle{apj} 
\bibliography{apj-jour,paper}

\clearpage
\begin{figure}
\includegraphics[angle=90,scale=.75]{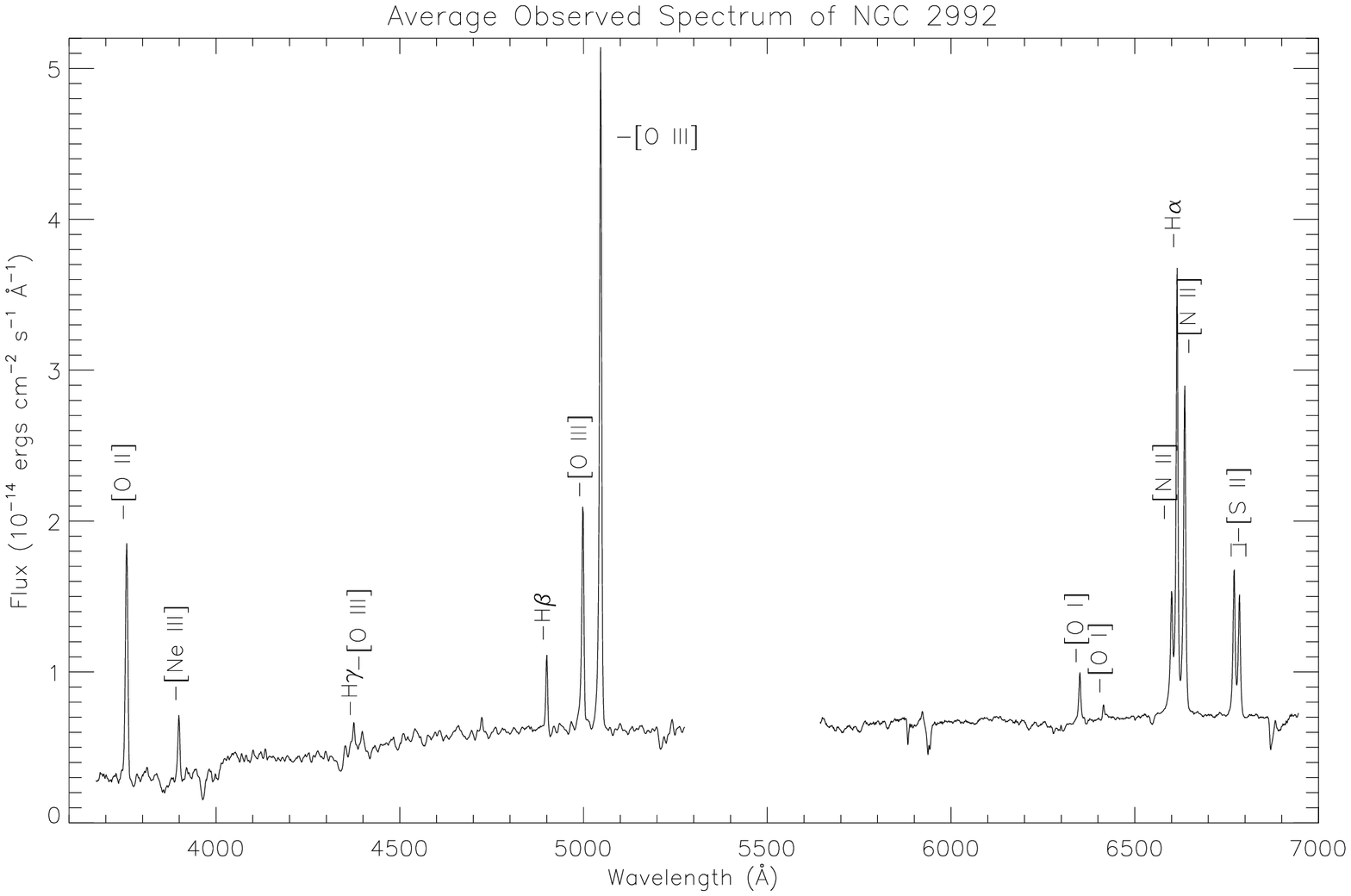}
\caption{Average observed spectrum of NGC 2992.
\label{fig1}}
\end{figure}
\clearpage
\begin{figure}
\includegraphics[angle=0,scale=.75]{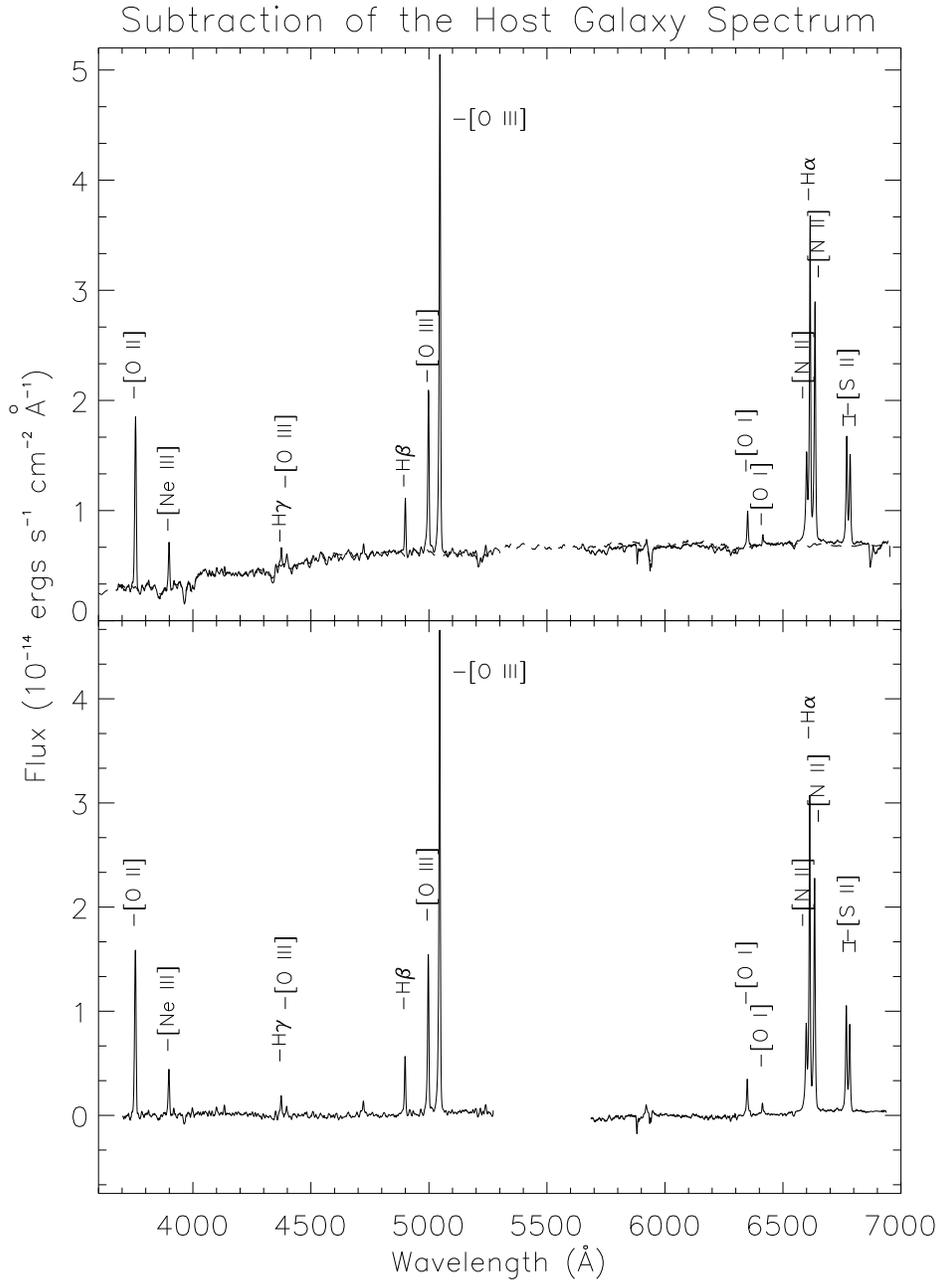}
\caption{Top, the average observed spectrum of NGC 2992 with overplot of host galaxy fit (dashed line). Bottom, observed spectrum after subtraction of host galaxy fit.
\label{fig2}}
\end{figure}
\clearpage
\begin{figure}
\includegraphics[angle=90,scale=.75]{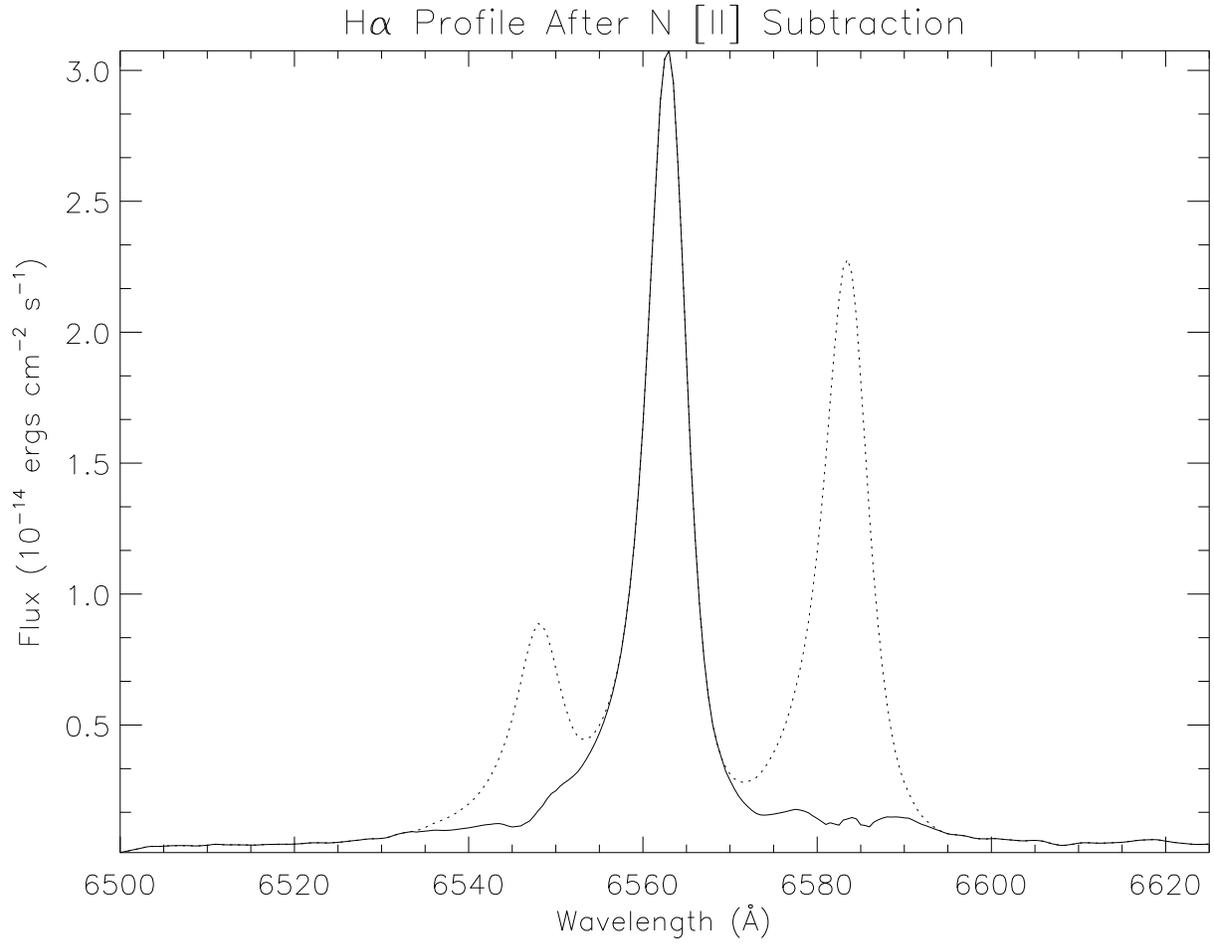}
\caption{The H$\alpha$ profile before (dotted line) and after (solid line) subtraction of the [N II] doublet. If there is a broad component, it is very weak.
\label{fig3}}
\end{figure}
\clearpage
\begin{figure}
\includegraphics[angle=0,scale=.75]{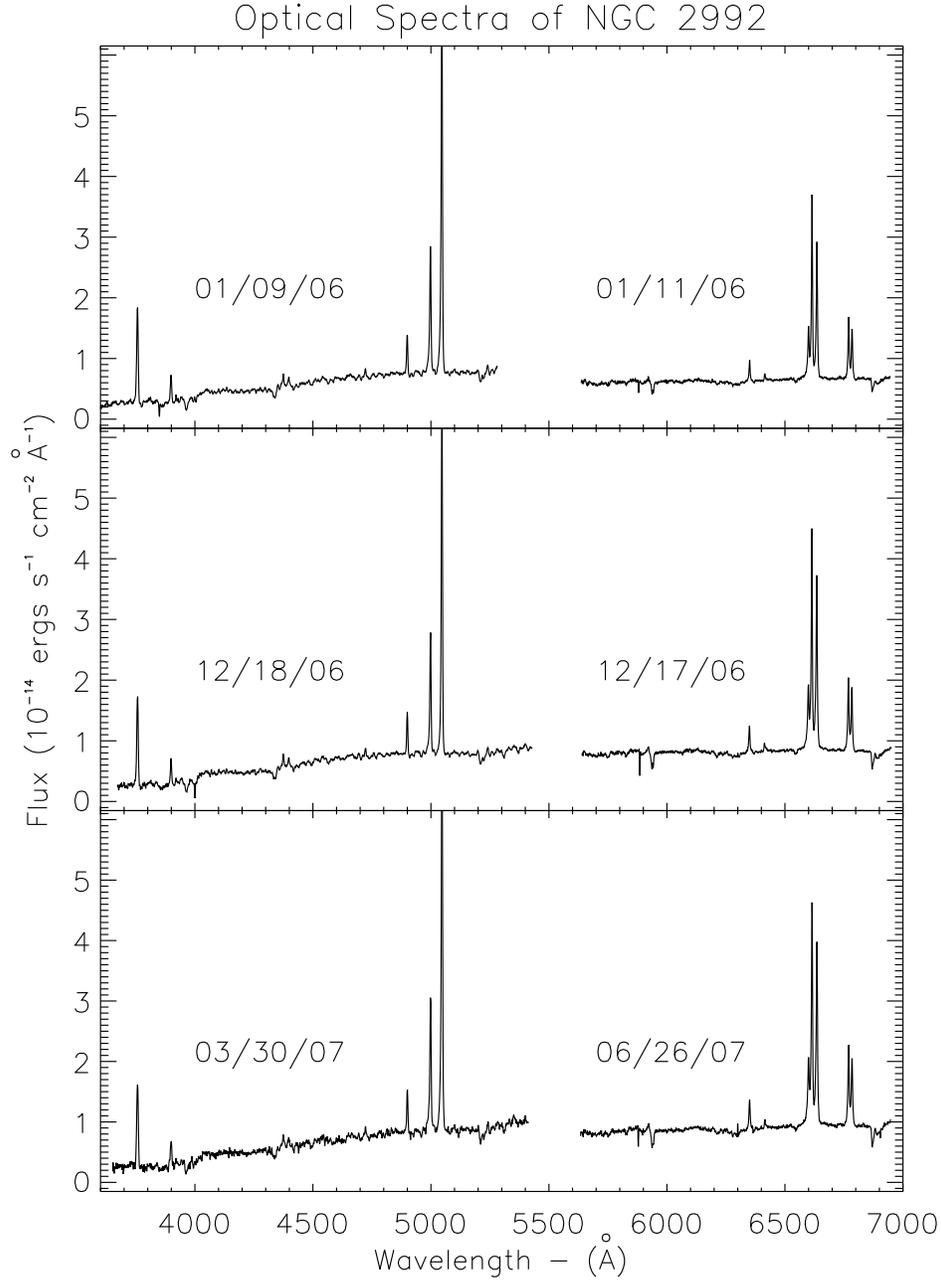}
\caption{Three representative blue and red spectra from the beginning, middle, and end of our monitoring campaign. The slight differences in absolute flux are likely due to seeing and/or non-photometric conditions, as noted in the text.
\label{fig4}}
\end{figure}
\clearpage
\begin{figure}
\includegraphics[angle=90,scale=.75]{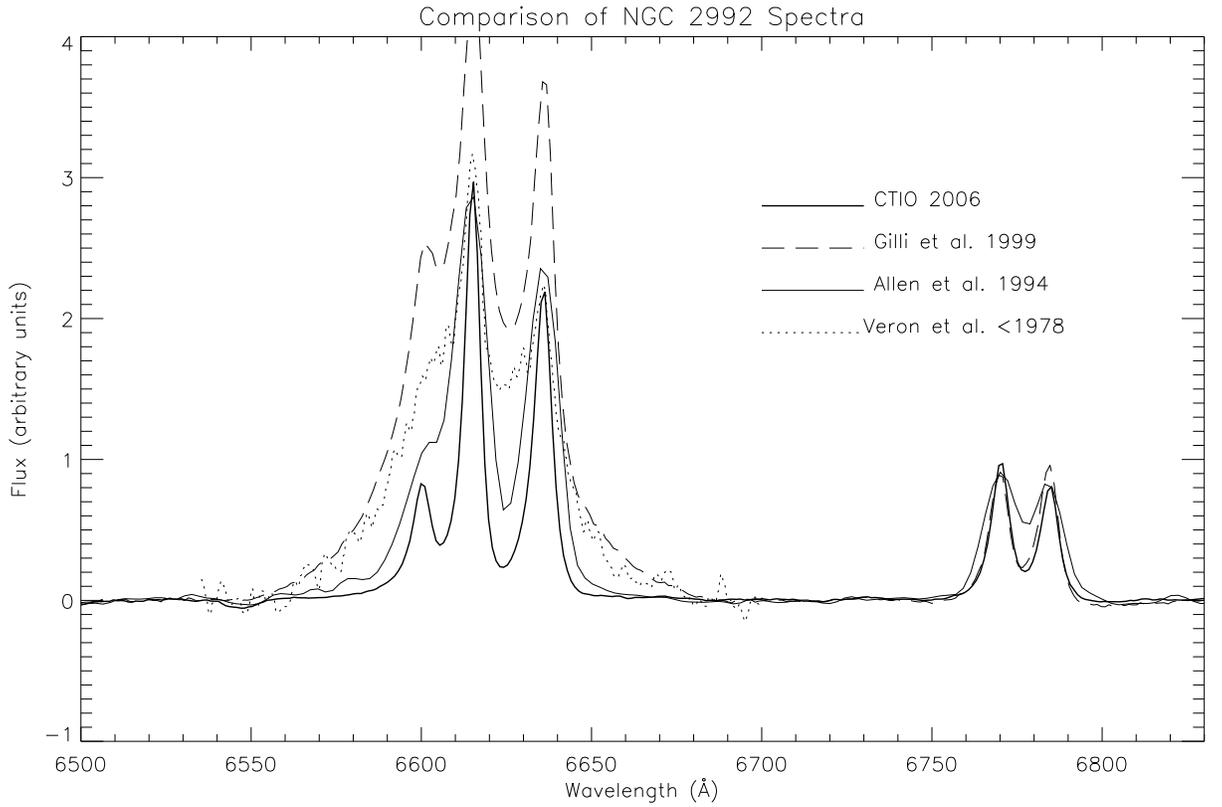}
\caption{The variation of NGC 2992's broad H$\alpha$ with time. For comparison purposes, we have subtracted the continuum and scaled the spectra in such a way that the narrow [S II] lines, which are expected to remain constant, match. NGC 2992 has dropped back into a low optical flux state, similar to that observed by Allen et al. in 1994, since it was last observed by Gilli et al. in 1999 
\citep[see][]{gil00}. \label{fig5}}
\end{figure}
\clearpage
\begin{figure}
\includegraphics[angle=90,scale=.75]{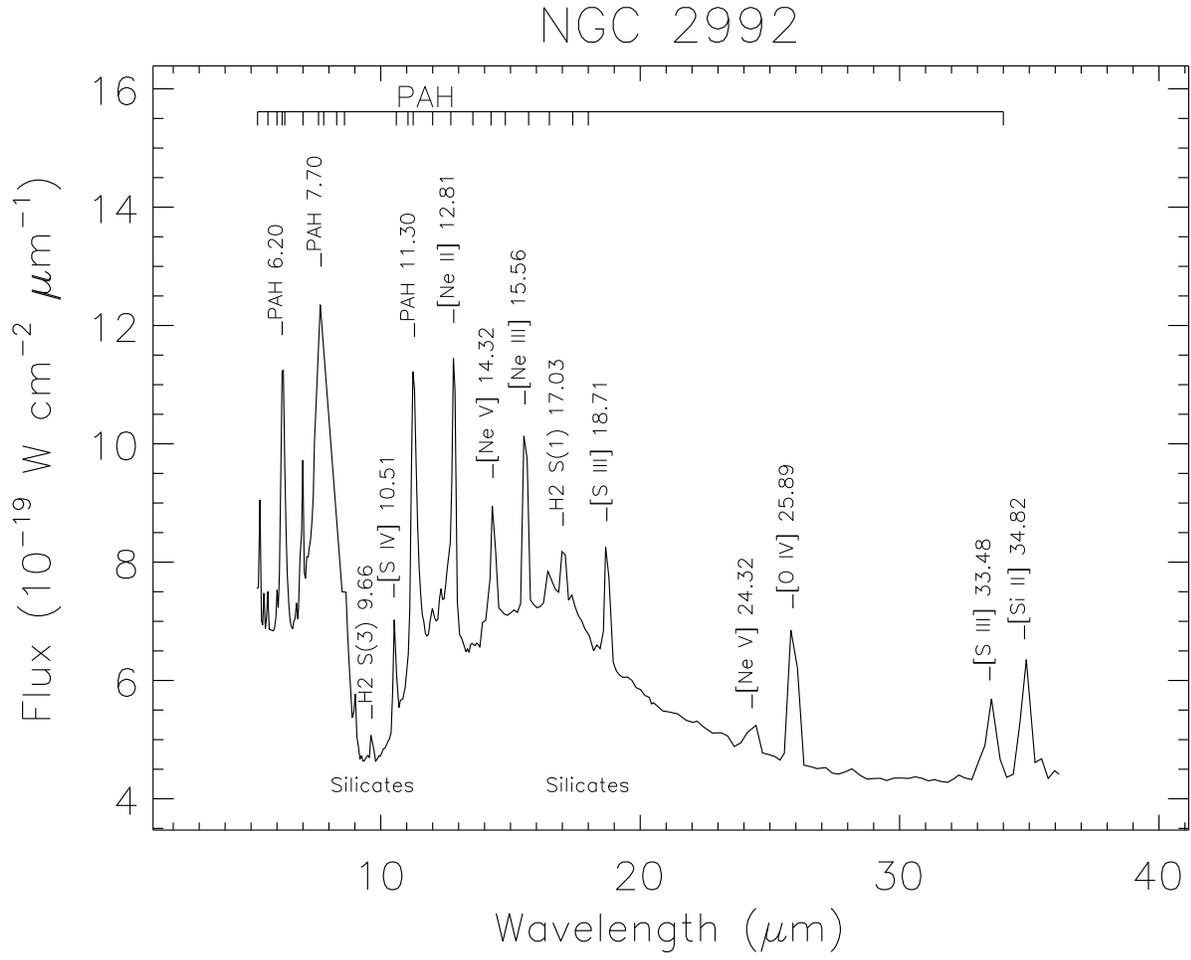}
\caption{Spitzer IRS spectrum of NGC 2992. \label{fig6}}
\end{figure}
\clearpage
\begin{figure}
\includegraphics[angle=0,scale=0.75]{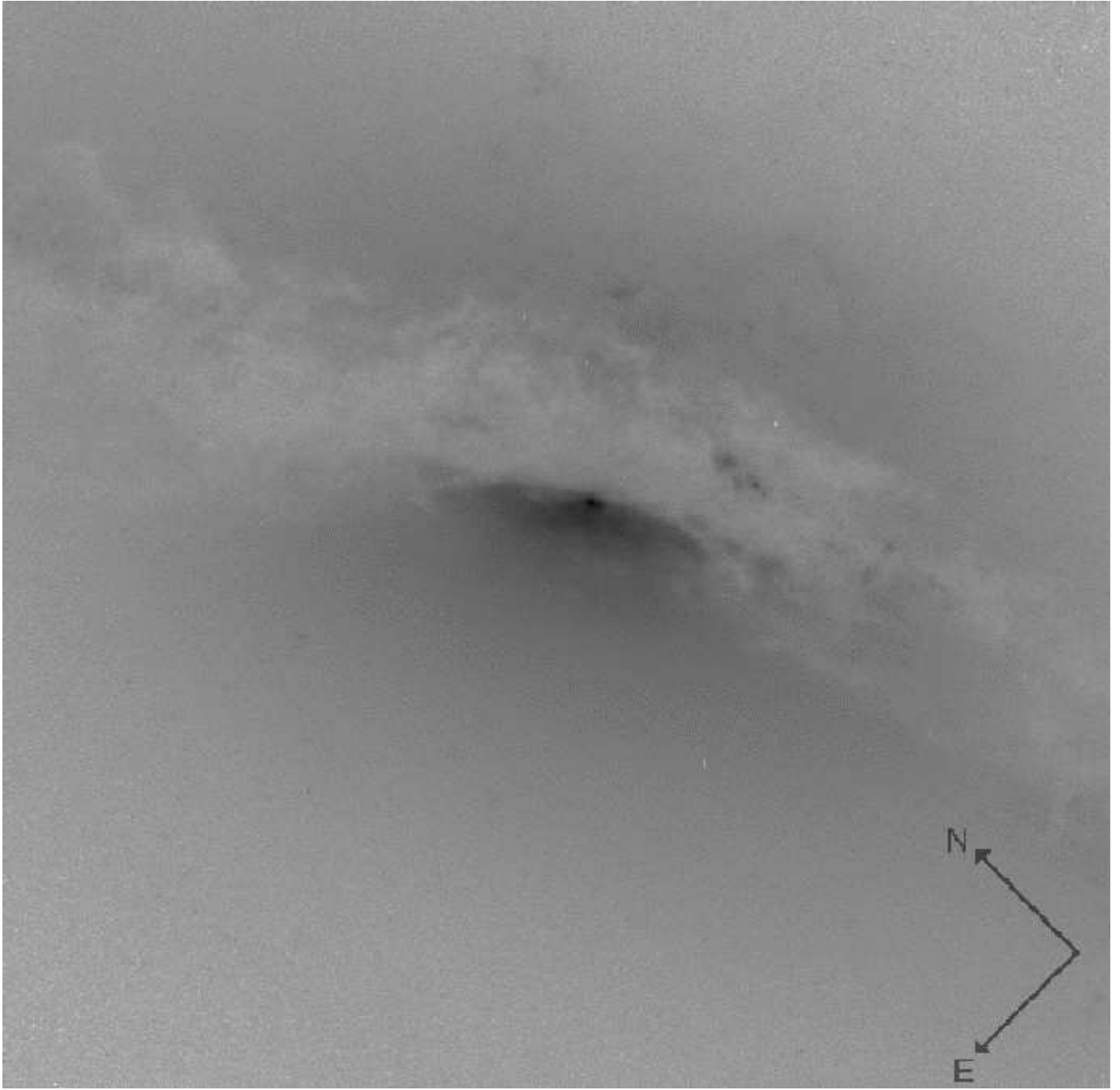}
\caption{Hubble WFPC2 image of NGC 2992 taken with the F606W filter, showing a large lane of dust crossing directly over the nucleus, inverted so emission is dark and absorption is light. This image is from the Space Telescope Science Institute's Multimission Archive. \label{fig7}}
\end{figure}

\clearpage
\begin{deluxetable}{lccc}
\tabletypesize{\scriptsize}
\tablewidth{0pt}
\tablecaption{Log of Observations}
\tablehead{
\colhead{Date} & \colhead{Wavelength Coverage} & \colhead{Exposure Time} & \colhead{Slit Width} \\
\colhead{$($U.T.$)$} & \colhead{$($\AA$)$} & \colhead{$($s$)$} & \colhead{$($arcsec$)$}
}
\startdata
2006 Jan. 9 & 3660-5440 & 1,200 & 2.0 \\
2006 Jan. 11 & 5650-6970 & 900 & 2.0 \\
2006 Feb. 2 & 5650-6970 & 900 & 1.5 \\
2006 Feb. 3 & 3660-5440 & 1,200 & 2.0 \\
2006 March 9 & 5650-6970 & 900 & 3.3 \\
2006 March 10 & 3660-5440 & 1,200 & 3.3 \\
2006 April 6 & 3660-5440 & 1,200 & 2.0 \\
2006 May 8 & 3660-5440 & 1,200 & 1.5 \\
2006 May 9 & 5650-6970 & 900 & 4.0 \\
2006 June 4 & 5650-6970 & 900 & 4.0 \\
2006 June 7 & 3660-5440 & 1,200 & 2.0 \\
2006 Nov. 29 & 3660-5440 & 1,200 & 2.0 \\
2006 Dec. 17 & 5650-6970 & 900 & 2.0 \\
2006 Dec. 18 & 3660-5440 & 1,200 & 2.0 \\
2007 Jan. 16 & 5650-6970 & 900 & 2.0 \\
2007 Jan. 17 & 3660-5440 & 1,200 & 2.0 \\
2007 Feb. 9 & 5650-6970 & 900 & 4.0 \\
2007 Feb. 8 & 3660-5440 & 1,200 & 4.0 \\
2007 April 3 & 5650-6970 & 900 & 2.0 \\
2007 March 30 & 3660-5440 & 1,200 & 2.0 \\
2007 June 21 & 3660-5440 & 1,200 & 4.0 \\
2007 June 26 & 5650-6970 & 900 & 2.0 \\
\enddata
\label{tbl1}
\end{deluxetable}

\clearpage
\begin{deluxetable}{lcc}
\tabletypesize{\scriptsize}
\tablecaption{Narrow-line ratios for NGC 2992 \label{tbl-2}}
\tablewidth{0pt}
\tablehead{
\colhead{Line} & \colhead{Observed Ratios to H$\beta$} & \colhead{Dereddened Ratios\tablenotemark{a}}
}
\startdata
$[$O II] 3727 & 3.86$\pm$0.48 & 7.09$\pm$0.17 \\ 
$[$Ne III] 3869 & 0.93$\pm$0.13 & 1.58$\pm$0.80 \\ 
H$\gamma$ 4340 & 0.85$\pm$0.17 & 1.16$\pm$0.17 \\ 
$[$O III] 4363 & 0.67$\pm$0.19 & 0.89$\pm$0.20 \\ 
H$\beta$ 4861\tablenotemark{b} & 1.00$\pm$0.17 & 1.00$\pm$0.17 \\ 
$[$O III] 4959 & 3.68$\pm$0.54 & 3.51$\pm$0.54 \\ 
$[$O III] 5007 & 10.27$\pm$1.27 &9.53$\pm$1.30 \\ 
$[$O I] 6300 & 0.77$\pm$0.17 & 0.39$\pm$0.15 \\ 
$[$O I] 6374 & 0.18$\pm$0.04 & 0.09$\pm$0.03 \\ 
$[$N II] 6548 & 2.21$\pm$0.29 & 1.03$\pm$0.28 \\ 
H$\alpha$ 6563 & 6.28$\pm$0.79 & 2.90$\pm$0.78 \\ 
$[$N II] 6583 & 5.24$\pm$0.65 & 2.40$\pm$0.64 \\ 
$[$S II] 6716 & 2.44$\pm$0.30 & 1.07$\pm$0.29 \\ 
$[$S II] 6731 & 1.93$\pm$0.24 & 0.84$\pm$0.23 \\ 
\enddata
\tablenotetext{a}{For $E(B-V)=0.71\pm0.11$ mag}
\tablenotetext{b}{The integrated absolute flux of the H$\beta$ line in this spectrum is 33.7$\pm$4.2 $\times10^{-15}$ ergs cm$^{-2}$ s$^{-1}$}
\label{tbl2}
\end{deluxetable}

\clearpage
\begin{deluxetable}{lcccr}
\tabletypesize{\scriptsize}
\tablecaption{Summary of Absorption Determinations \label{tbl-3}}
\tablewidth{0pt}
\tablehead{
\colhead{Region} & \colhead{$E(B-V)$} & \colhead{$N_{H}$ $(10^{21}$cm$^{-2})$} & \colhead{Date of Obs.} & \colhead{Reference}
}
\startdata
X-ray emission region & -- & 7.99$\pm$0.6 & 2005 &\citet{yaq07}\\ 
X-ray emission region & -- & 9.0$\pm$0.3 & 1998 &\citet{gil00}\\ 
X-ray emission region & -- & 14.0$\pm$5.0 & 1997 & \citet{gil00}\\ 
BLR & 0.71$\pm$0.19 & 3.7$\pm$0.99 & 1999 &\citet{gil00}\\ 
NLR & 0.71$\pm$0.11 & 3.7$\pm$0.57 & 2006-2007 & This work.\\ 
NLR & 0.65$\pm$0.19 & 3.4$\pm$0.99 & 1999 &\citet{gil00}\\ 
External galaxy & 2.1$\pm$0.20 & 11.0$\pm$1.0 & 2005 & This work.\\ 
\enddata
\label{tbl3}
\end{deluxetable}

\end{document}